\documentclass[aps,prl,twocolumn,superscriptaddress,showkeys]{revtex4-1}

\usepackage[T1]{fontenc}
\usepackage{graphicx}  
\usepackage{dcolumn}   
\usepackage{bm}        
\usepackage{amssymb}   
\usepackage{amsmath}
\usepackage{float}
\usepackage{color}
\usepackage{latexsym}
\usepackage{amsthm}
\usepackage{epsfig}
\usepackage{soul}
\usepackage{hyperref}
\hypersetup{colorlinks=true, citecolor=blue}

\begin{document}

\preprint{APS/123-QED}

\title{Additional transition line in jammed asymmetric bidisperse granular packings}

\author{Juan C. Petit}
\affiliation{Institut f\"{u}r Materialphysik im Weltraum, Deutsches Zentrum f\"{u}r Luft- und Raumfahrt (DLR), 51170 K\"{o}ln, Germany }%
\author{Nishant Kumar}%
\affiliation{Institut f\"{u}r Materialphysik im Weltraum, Deutsches Zentrum f\"{u}r Luft- und Raumfahrt (DLR), 51170 K\"{o}ln, Germany }%
\author{Stefan Luding}%
\affiliation{Multi-Scale Mechanics (MSM), Faculty of Engineering Technology, MESA+,
University of Twente, Enschede, The Netherlands }%
\author{Matthias Sperl}%
\affiliation{Institut f\"{u}r Materialphysik im Weltraum, Deutsches Zentrum f\"{u}r Luft- und Raumfahrt (DLR), 51170 K\"{o}ln, Germany }%
\affiliation{Institut f\"{u}r Theoretische Physik, Universit\"{a}t zu K\"{o}ln, 50937 K\"{o}ln, Germany}%

\date{\today}

\begin{abstract}

We present numerical evidence for an additional discontinuous transition
inside the jammed regime for an asymmetric bidisperse granular packing
upon compression. This additional transition line separates jammed states
with networks of predominantly large particles from jammed networks formed
by both large and small particles, and the transition is indicated by a
discontinuity in the number of particles contributing to the jammed
network. The additional transition line emerges from the curves of jamming
transitions and terminates in an end-point where the discontinuity
vanishes. The additional line is starting at a size ratio around $\delta =
0.22$ and grows longer for smaller $\delta$. For $\delta \to 0$, 
the additional transition line approaches a
limit that can be derived analytically. The observed jamming scenarios are
reminiscent of glass-glass transitions found in colloidal glasses.

\end{abstract}

\maketitle

Jamming governs the transition to rigidity of athermal amorphous systems. {Granular matter is one example showing its jamming 
density, $\phi_J$, at a packing fraction marked by a discontinuous jump in the contact number}
\cite{zhang2005jamming, aharonov1999rigidity, van2009jamming, behringer2018physics, o2003jamming, majmudar2007jamming}. For frictionless 
monodisperse packings, such a value is close to $\phi_{J} \approx 0.64$ in 3D \cite{donev2004jamming, o2003jamming}. For bidisperse packings, 
introduced to consider a higher degree of complexity in the system and to suppress crystallization, $\phi_J$ can be tuned to higher values by varying the size ratio, $\delta$, and volume concentration of small particles, $X_{\mathrm S}$, 
\cite{prasad2017subjamming,hopkins2011phase, biazzo2009theory, pillitteri2019jamming, kumar2016memory}. This is relevant for industrial 
processes since mechanical properties of bidisperse packings such as bulk 
modulus and wave speed can be controlled \cite{kumar2016tuning}.

Although the dependence of the jamming density on $\delta$ and $X_{\mathrm S}$ has been studied previously 
\cite{prasad2017subjamming,hopkins2011phase,biazzo2009theory, pillitteri2019jamming,  
kumar2016memory}, a better understanding of jammed states in highly asymmetric bidisperse mixtures for extremely low $X_{\mathrm S}$ 
is intended. For example, previous works have assumed that the jammed structure of a bidisperse packing is formed by the equal contribution of 
both large and small particles since both species 
jam simultaneously at $\phi_J$ \cite{hopkins2011phase, biazzo2009theory, pillitteri2019jamming}. This is true above 
certain values of $\delta$ and $X_{\mathrm S}$, since then the particle sizes are similar enough and concentration of small particles is 
high enough that both species follow the same behavior. However, for lower values of $\delta$ and $X_{\mathrm S}$, each 
component behaves differently when approaching jamming, which suggests a decoupling in the jamming in the binary system. Indeed, it 
has been recently evidenced in Ref.~\cite{prasad2017subjamming} that the jammed structure of an extremely asymmetric bidisperse 
system evolves from a small-sphere-rich to a small-sphere-poor structure. Such behavior is marked by an abrupt decay in the number of 
small particles contributing to the jammed stucture at a specific $X_{\mathrm S}$, while the rest of small particles remain 
without contacts.

This indicates that in principle, two different pathways to jamming need to be distinguished for small $\delta$: one driven by
predominantly large particles, and one driven by both together. Our aim in this letter is to demonstrate that indeed, the distinction can be made 
rigorous by an analy-sis of the partial contact numbers between particles of different species. It leads in particular to the identification 
of a new transition between \emph{different jammed states} for $X_{\mathrm S} < X_{\mathrm S}^{*}$.

To investigate the emergence of this new transition, we used MercuryDPM to perform 3D Discrete Element 
Method (DEM) simulations, suitable to study granular systems 
\cite{cundall1979discrete, thornton2010quasistatic,thornton2010evolution}. Here, we consider bidisperse 
packings formed by $N = 6000$ particles, where a number of large, $N_{\mathrm L}$, and 
small, $N_{\mathrm S}$, particles with radius $r_{\mathrm L}$ and $r_{\mathrm S}$ are considered, respectively. We characterize 
each packing by the size ratio, $\delta = r_{\mathrm S}/r_{\mathrm L}$, and the volume concentration of small particles, 
$X_{\mathrm S} = N_{\mathrm S} \delta^{3} / (N_{\mathrm L} + N_{\mathrm S} \delta^{3})$. 
We use the linear normal contact force model given as 
${\it \bf f}^n_{ij}=f^n_{ij} \hat{\bf n} = ( k_{n} \alpha_c + \gamma_{n} \dot\alpha_c) \hat{\bf n}$ 
\cite{cundall1979discrete, luding2008cohesive, kumar2016tuning}, where $k_{n}$ is the contact spring stiffness, $\gamma_{n}$ 
is the contact damping coefficient, $\alpha_c$ is the contact overlap and $\dot\alpha_c$ is the relative velocity in the normal 
direction $\hat{\bf n}$. An artificial background dissipation force, ${\it \bf f}_b=-\gamma_b {\bf v}_i$, proportional to the 
velocity ${\bf v}_i$ of particle $i$ is added, resembling the damping due to a background medium.
We fix $r_{\mathrm L} = 1.5\,{\rm mm}$ and vary $r_{\mathrm S} \in [0.255, 1.5]\,{\rm mm}$, such that $\delta \in [0.15, 1]$.
Both small/large particles have the same interacting properties, i.e., 
$\rho = 2000 \,\rm{kg/m^{3}}$, $k_{n} = 10^{5} \,\rm{kg/s^{2}}$ and $\gamma_{n} = \gamma_b = 1 \,\rm{kg/s}$, which represent glass beads since
they are most spherical and easily available material in experiments. The contact duration, $t_{c}$, and restitution coefficient, $e$, 
depend on the particle sizes. 
The fastest response time scale corresponding to the interaction between smallest particles according to the lowest value of $\delta$ is
$t_{c} = 0.013 \,\rm{\mu s}$ and $e = 0.996$. As usual in DEM, the time step was chosen to be 50 times smaller than the shortest time
scale $t_{c}$. The contact and background dissipation value is used to reduce computational time 
during relaxation. We restrict ourselves to isotropic deformation and the linear contact model 
without any friction between particles \citep{kumar2014macroscopic}. Thus, we exclude all 
the non-linearities present in the system due to contact models and analyze the effect of size and concentration 
at jamming. 

\begin{figure}[t]
\centering \includegraphics[scale=0.34]{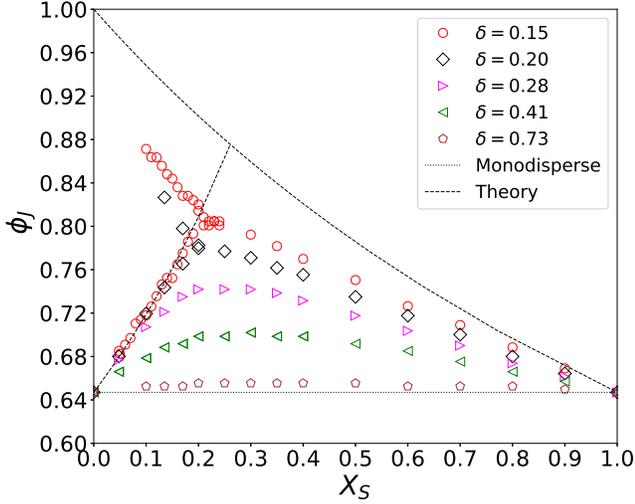}
\caption{Jamming density, $\phi_{J}$, as a function of the volume concentration of small particles, $X_{\mathrm 
S}$, for different values of the size ratio, $\delta$. The extreme $X_{\mathrm S}$ values (0 and 1) 
correspond to monodisperse systems where $\phi_{J}=\phi_\text{RCP} \approx 0.64$
indicated by the dashed horizontal line. Solid lines represents the analytical result of 
Eq.~(\ref{ecu1}) \cite{furnas1931grading}.
\label{fig1}
}
\end{figure}

Each bidisperse packing corresponding to a combination of ($\delta$, $X_{\mathrm S}$) is created and 
further compressed using a unique, well defined protocol \citep{goncu2010constitutive}. 
For each packing, configurations near (below and above) their jamming density are picked from the decompression branch. This 
is allowing packings to dissipate their kinetic energy and reach quickly unjammed, non-overlapping packings 
\footnote{Configurations from the decompression branch are more reliable since they are much less sensitive to 
the protocol and rate of deformation during preparation, see \cite{goncu2010constitutive}, but the 
$\phi_J$ does then depend on the maximum $\phi_{\mathrm{max}}$ \cite{kumar2016memory}.}. Therefore, 
$\phi_J$ for each configuration is determined at the point where the partial mean contact number exhibits a sharp
drop. Note that this protocol allows to identify multiple jamming densities occurring at different maximum packing 
fractions \cite{kumar2016memory}. The values of $\phi_J$ as a function of $X_{\mathrm S}$ for different $\delta$ are 
shown in Fig.~\ref{fig1}.

\begin{figure}[t]

    \centering \includegraphics[scale=0.45]{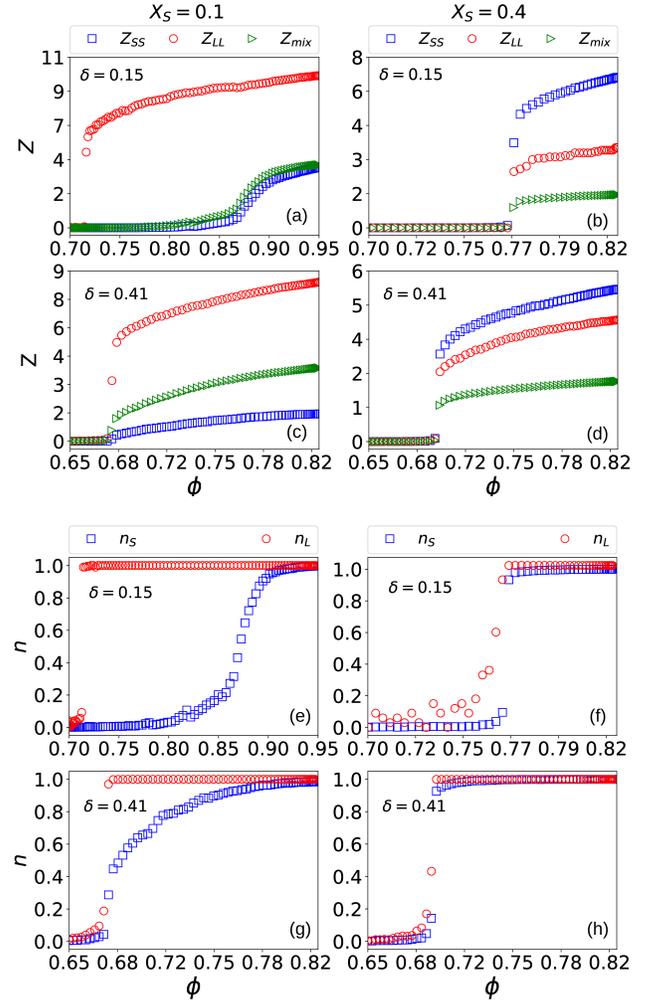}

 \caption{(a-d) Mean contact number, $Z$, for each type of contact as a function of $\phi$ for four tuples of $\delta$ and $X_{\mathrm S}$.
 $Z_{\mathrm{nm}}$  corresponds to the sum of contacts between particles $\rm{n}$ and $\rm{m}$ divided by total number of $N_{\rm n}$ particles, with 
 $\rm{n,m} \in [L,S]$. The analyzed data were along the decompression branch. $Z_{\rm mix}$ is defined as $Z_{\rm mix} = (Z_{\mathrm{LS}} + Z_{\mathrm{SL}})/N$. 
 (e-h) Fractions of large, $n_{\mathrm L}$, and small,  $n_{\mathrm S}$, particles contributing to the force network as a function of the packing fraction, 
 $\phi$, for two sets of ($\delta$, $X_{\mathrm S}$).}
  \label{fig2}
\end{figure}

For a particular $\delta$, the jamming density increases with $X_{\mathrm S}$ up to a maximum value occurring at $X_{\mathrm S}^{*} \approx 0.21$, 
then decreases for larger values. Along the increasing transition line, jamming is driven by the force network created by the large particles 
since most of the small ones remain without or only few contacts in the cages formed by the large ones. This is confirmed 
when looking at Fig.~\ref{fig2} (a,c), where mainly large particles are carrying the load in the jammed state. On the other hand, on the decreasing branch 
of $\phi_J$ for high $X_{\mathrm S}$ small and large particles jam simultaneously at the same density, see Fig.~\ref{fig2} (b,d). However, for lower 
values of $\delta$ and $X_{\mathrm S}$ a decoupling in the mean contact number, $Z$, between large and small particles is observed,  see Fig.~\ref{fig2} (a).
Large particles first jam at lower $\phi$ while small ones remain without contacts. Making the packing denser, small particles exhibit an apparent jump in $Z_{\mathrm{SS}}$ and 
$Z_{\mathrm{mix}}$, see Fig.~\ref{fig2} (a), indicating that a fraction of them start contributing in a discontinuous fashion to
the already jammed structure of large ones. This clearly suggests that the line of jammed states encountered at high $X_{\mathrm S}$ need 
to be extended at lower $X_{\mathrm S}$ values as well in this case.

\begin{figure}[t]

\centering \includegraphics[scale=0.34]{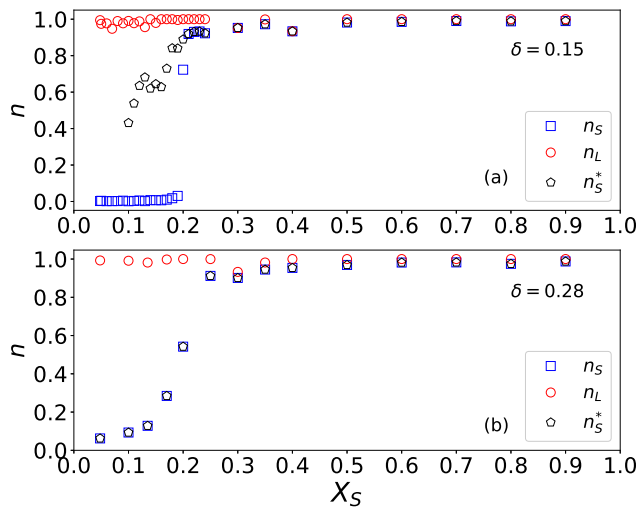}
 \caption{Fraction of small, $n_{\mathrm{S}}$, and large particles, $n_{\mathrm{L}}$ contributing 
to the jammed structure as a function of the concentratin $X_{\mathrm S}$ of
small particles for 
 (a) $\delta = 0.15$ and (b) $\delta = 0.28$. $n_{\mathrm{S}}^{*}$ 
represents the fraction of small particles along the 
additional
 line.}
  \label{fig3}
\end{figure}

Since it is difficult to identify where $Z_{\mathrm{SS}}$ jumps for smaller $X_{\mathrm S}$, we have quantified the fraction of large, $n_{\mathrm L}$, and small 
particles, $n_{\mathrm S}$, contributing to the force network as a function of $\phi$ giving a clearer jump compared to $Z$. 
A clear decoupling between $n_{\mathrm{L}}$ and $n_{\mathrm{S}}$ at lower $\delta$ and $X_{\mathrm S}$ can be seen in Fig.~\ref{fig2} (e), 
while for higher values of $\delta$, both size of particles contribute simultaneously at the jammed structure. Such decoupling indicates that a big amount of 
small particles are jammed discontinuously at higher densities, which resembles a similar behavior done by large particles at low densities. To extract the 
precise value of the jamming density where $n_{\mathrm S}$ jumps for low $X_{\mathrm S}$, we computed the derivative of $\partial n_{\mathrm S}/ \partial \phi$ 
and found a specific density, which corresponds to $\phi_J$ of the system. Fig.~\ref{fig1} 
displays such $\phi_{J}^{c}$ values for $X_{\mathrm S} < X_{\mathrm S}^{*} \approx 0.21$ indicated by the increasing line of densities for smaller $X_{\mathrm S}$. 
Such a line clearly extends the transition where both size particles are jammed for low range of $X_{\mathrm S}$, thus 
introducing a more complete jamming diagram for bidisperse packings.

The result shown in Fig.~\ref{fig1} is quite similar to those previously found 
in experiments using glass beads \cite{pillitteri2019jamming} and in 3D simulations \cite{hopkins2011phase, biazzo2009theory}, 
but they did not investigate the impact of the small particles contribution on the jammed 
structure for low $X_{\mathrm S}$ values and no additional line was shown. Recent work 
\cite{prasad2017subjamming}
which has studied the statistics of the small particles in the jammed structure of bidisperse system in more detail, showed that 
indeed the fraction of small particles contributing to the jammed structure decays to zero for $X_{\mathrm S} < X_{\mathrm S}^{*}$, 
which becomes discontinuous as $\delta$ decreases. This allows to make a distinction between two states, one where small particles
contribute to the jammed network jointly with large ones ($X_{\mathrm S} \geq X_{\mathrm S}^{*}$) and the second one where only large 
particles contribute ($X_{\mathrm S} < X_{\mathrm S}^{*}$). We have found that besides these two results, there is an additional line emerging at 
$X_{\mathrm S} \approx X_{\mathrm S}^{*}$. Such line represents a jammed state, $\phi_{J}$, extending at higher 
densities. In addition, we found that such line starts emerging for $\delta < 0.22$ where its end-point terminates at lower 
$X_{\mathrm S}$ as $\delta$ decreases.

The additional transition line obtained is a result of the decoupling of the jamming transition 
between large and small particles 
for low $X_{\mathrm S}$ and low $\delta$. Such a scenario can be understood by a model 
introduced by Furnas almost a century ago \cite{furnas1931grading} to predict the highest density of aggregates entering in
the manufacture of mortar and concrete. This model suggests that if the size ratio of the particles 
are extreme ($\delta \to 0$), $\phi_J$ can decouple in two limits sharing a common point at $\hat{X}^{*}_{\mathrm S}$. One limit 
considers an approximation where large particles dominate the jammed structure while small particles are not taken 
into account since the number of them are not enough to play a role ($0 \leq X_{\mathrm S} < \hat{X}^{*}_{\mathrm S}$). In the 
second limit, both large and small particles participate in the jammed structure ($0 \leq X_{\mathrm S} \leq 1$). In this 
case, the number of small particles are high enough to drive few large particles to the jammed state. Both limits
are written as 

\begin{equation}
\lim_{\delta \to 0} \phi_{J}(X_{\mathrm S}) =
  \begin{cases}
    \frac{\phi_{\mathrm{RCP}}}{(1 - X_{\mathrm S})} \,\, & \text{if } 0 \leq X_{\mathrm S} < \hat{X}^{*}_{\mathrm S},\\
     \frac{\phi_{\mathrm{RCP}}}{[\phi_{\mathrm{RCP}} + (1 - \phi_{\mathrm{RCP}})\,X_{\mathrm S}]}  &  \text{if } 0 \leq X_{\mathrm S} \leq 1,
  \end{cases}
\label{ecu1}
\end{equation}

and are displayed in Fig.~\ref{fig1}, which represent the maximum densities that a highly asymmetric binary mixture can achieve  
while changing the concentration of small particles. Note that this model describes the trend of the data, fitting those values 
for low $X_{\mathrm S}$ which correspond to the single jammed state. It shows a maximum density of $\phi_J(\hat{X}^{*}_{\mathrm S}) \approx 0.87$ 
at $\hat{X}^{*}_{\mathrm S}=(1-\phi_\text{RCP})/(2-\phi_\text{RCP})\approx0.26$, which is in reasonable agreement with 
the value obtained here for $X^{c}_{\mathrm S}$ at finite $\delta$. Interestingly, the model shows 
an additional transition line right where both limits meet, ending at a density of unity. However, previous works using 
the Furnas model disregarded the physics behind such additional transition line. In 
Fig.~\ref{fig1}, the additional line given by our simulation data follows
qualitatively the theoretical prediction, ending at $X_{\mathrm S}^{\circ} = 0.1$ for the lowest $\delta$. 
The transition line obtained from simulation stops at this value since for $X_{\mathrm S} < X_{\mathrm S}^{\circ}$ no jump in 
$n_{\mathrm{S}}$ and $Z_{\mathrm{SS}}$ was found, instead these quantities increase continuously in this region, not showing features 
of a jump transition. This allows us to argue that the additional transition line terminates in an 
end-point at some finite $X_{\mathrm S}^{\circ}$, 
which depends on $\delta$.

The evolution of the fractions of small and large particles that contribute to the jammed structure, 
$n_{\mathrm{S}}$ and $n_{\mathrm{L}}$, along the jamming lines elucidates the different nature of the 
transitions, i.e., by discussing the jump heights in these quantities as the jamming transition is crossed. 
These ratios are shown in Fig.~\ref{fig3} for two representative values of $\delta$. Here, we extracted 
$n_{\mathrm{S}}$ and $n_{\mathrm{L}}$ from the points where the large-particle contribution jumps, and 
$n_{\mathrm{S}}^*$ from those where the small-particle contribution jumps. The existence of the 
additional transition line is now manifested in the cases where $n_{\mathrm{S}}^*$ splits from 
$n_{\mathrm{S}}$, see Fig.~\ref{fig3} (a) at low $X_{\mathrm S}$. Close to the crossing point where the 
additional transition line emerges, the difference between $n_{\mathrm{S}}$ and $n_{\mathrm{S}}^*$ is largest.
For lower $X_{\mathrm S}$, the second jump $n_{\mathrm{S}}^*$ has to be compared with the value $n_{\mathrm{S}}$ 
that is increasing regularly with packing fraction after the first transition: Once $n_{\mathrm{S}}$ and 
$n_{\mathrm{S}}^*$ become of equal value, no more second jump can be identified the endpoint of the additional 
transition line reached. Technically, such a cessation 
of a second jump is more difficult to identify precisely than a clearly developed second jump close to the 
crossing point which explains the fluctuations in Fig.~\ref{fig3}.
The sharp rise in $n_{\mathrm{S}}$ around $X_{\mathrm S}^{*} \approx 0.21$ has been noted before \cite{prasad2017subjamming} and 
connected with a ``sub-jamming'' transition. It can be seen as a natural consequence from the crossing of a line with a finite jump in 
$n_{\mathrm{L}}$ and a line with a finite jump in $n_{\mathrm{S}}^*$. For $\delta = 0.28$, Fig.~\ref{fig3} (b), $n_{\mathrm{S}}^*$ and 
$n_{\mathrm{S}}$ merged into one line at lower $X_{\mathrm S}$ suggesting that no additional transition line is found, see Fig.~\ref{fig1}.

To summarize, we have shown that the jamming diagram in bidisperse packings is enriched by an additional transition.
This transition appears for $X_{\mathrm S} < X_{\mathrm S}^{*}$ when small particles 
get in contact with the jammed structure already formed by large particles. The data for the lowest size asymmetry, 
$\delta = 0.15$, is well described analitically by the Furnas model which predicts a similar 
additional line.

The results presented in Fig.~\ref{fig1} demonstrate an interes\-ting connection to the glass-glass transition phenomeno\-logy 
experienced by a binary colloidal suspensions previously found in \cite{voigtmann2011multiple}. The emergence of such a transition was 
found through a bifurcation line when the whole range of $\delta$ and $X_{\mathrm S}$ was explored. Similarly, small species become 
arrested when the system is becoming denser, thus separating two glassy states in the system. This clearly shows a surprising relation 
between a bidisperse granular system and a bidisperse colloidal system approaching the jamming density and the glass transition, 
respectively.

{ The extension of the jamming line that we found here gives a mathematically rigorous definition of the ``sub-jamming'' transition,
a term introduced in Ref.~\cite{prasad2017subjamming}. There, the crossing point $(\delta_c,X_{\mathrm S}^*)$ has been interpreted as showing the
hallmarks of a critical end-point akin to those of phase transitions. However, as it is evident here, this point is rather a \emph{crossing}
point of two bifurcation lines, forming part of a more general bifurcation diagram. Mathematically speaking, both jamming lines that
cross here can continue until their respective end-points, which need not coincide with $(\delta_c,X_{\mathrm 
S}^*)$. This additional transition line signals sub-jamming, i.e., small particles are 
inside the interstices of an already jammed structure of large particles. Increasing $\delta$, the 
additional transition line shrinks, so that at 
some critical $\delta_c$, the end-point and crossing-points coincide, $X_{\mathrm S}^{\circ}(\delta_c)=X_{\mathrm S}^*$. This (a point of ``higher order
bifurcation'' \cite{Arnold}) will mark the critical point whose signature was discussed in Ref.~\cite{prasad2017subjamming}.}

We thank T. Kranz, P. Born, L. Elizondo-Aguilera and Th. Voigtmann for helpful discussions about the results. 
This work was supported by the German Academic Exchage Service (DAAD).

\bibliography{Ref}

\bibliographystyle{apsrev4-1}

\end{document}